# Excitonic and valleytronic signatures of correlated states at fractional fillings of a moiré superlattice

Erfu Liu[1], Takashi Taniguchi[2], Kenji Watanabe[3], Nathaniel M. Gabor[1,4], Yong-Tao Cui[1], Chun Hung Lui[1*]

[1] Department of Physics and Astronomy, University of California, Riverside, CA 92521, USA.

[2] International Center for Materials Nanoarchitectonics (WPI-MANA), National Institute for Materials Science, 1-1 Namiki Tsukuba, Ibaraki 305-0044, Japan.

[3] Research Center for Functional Materials, National Institute for Materials Science, 1-1 Namiki, Tsukuba 305-0044, Japan.

[4] Canadian Institute for Advanced Research, MaRS Centre West Tower, 661 University Avenue, Toronto, Ontario ON M5G 1M1, Canada.

[*]Corresponding author. Emails: joshua.lui@ucr.edu.

**Abstract: Moiré superlattices are excellent platforms to realize strongly correlated quantum phenomena, such as Mott insulation and superconductivity. In particular, recent research has revealed stripe phases and generalized Wigner crystals at fractional fillings of moiré superlattices. But these experiments have not focused on the influence of electronic crystallization on the excitonic and valleytronic properties of the superlattices. Here we report excitonic and valleytronic signatures of correlated states at fractional fillings in a $WSe_2/WS_2$ moiré superlattice. We observe reflection spectral modulation of three intralayer moiré excitons at filling factors $v$ = 1/3 and 2/3. We also observe luminescence spectral modulation of interlayer trions at around a dozen fractional filling factors, including $v$ = -3/2, 1/4, 1/3, 2/5, 2/3, 6/7, 5/3. In addition, the valley polarization of interlayer trions is noticeably suppressed at some fractional fillings. These results demonstrate a new regime of light-matter interactions, in which electron crystallization significantly modulates the absorption, emission, and valley dynamics of the excitonic states in a moiré superlattice.**

The recent development of moiré superlattices formed by van der Waals materials opens a new path to study strongly correlated quantum phenomena, including Mott insulation, superconductivity, and correlated magnetic phases[1-15]. In particular, moiré superlattices formed by two monolayers of transition metal dichalcogenide (TMD) can exhibit correlated insulating states at fractional fillings, which have been interpreted as crystalline electron phases including the stripe phases and generalized Wigner crystals[7, 16-22]. In these semiconducting two-dimensional (2D) superlattices, the Coulomb interactions between carriers are markedly enhanced by the 2D confinement, yet the carrier kinetic energy is strongly suppressed by the large carrier effective mass and slow hopping between moiré cells. As a result, electrons tend to distribute orderly and periodically to reduce their total energy, leading to the formation of crystalline electron phases in analogy to the atomic crystals. Recent research has reported evidence of stripe phases and generalized Wigner crystals at fractional fillings of $WSe_2/WS_2$ heterobilayer moiré superlattices (Fig. 1) by probing their dielectric, conduction, and optical response[7, 16-19]. While the recent work provides evidence for moiré electron crystals, there is still a lack of studies on how these correlated phases affect the excitonic and valleytronic properties of the materials. Moiré superlattices formed by valley semiconductors (e.g. $MoS_2$, $WSe_2$, and $WS_2$) are known to host robust excitonic states with distinctive valleytronic properties, which hold promises for novel excitonic and valleytronic applications[23-32]. The coupling between these excitonic properties and the crystalline electron phases can enable entirely new methods to control the physical properties of van der Waals heterostructures. Therefore, it is intriguing to explore excitonic phenomena in the regime of electron crystallization.

In this Letter, we report the observation of excitonic and valleytronic signatures of correlated states at fractional fillings in a $WSe_2/WS_2$ moiré superlattice. We observe significant modulation of reflectance contrast of three $WSe_2$ intralayer moiré excitons at fractional fillings $v = 1/3$ and $2/3$. The two high-lying moiré excitons exhibit much higher sensitivity to the fractional correlated phases than the lowest-lying moiré exciton due to their higher mobility and different wavefunction structure. We also observe abrupt changes of the photoluminescence (PL) intensity and photon energy of the interlayer trions at a series of fractional fillings, including $v = -3/2, 1/4, 1/3, 2/5, 2/3, 6/7, 5/3$. Remarkably, we also find that the valley polarization of interlayer trions decreases noticeably at some fractional fillings. Therefore, the absorption, emission, and valley dynamics of the excitonic states all change significantly at fractional fillings. Our findings reveal that the emergent electron crystalline phases at fractional fillings, with their insulating nature, can effectively modulate the exciton and trion binding, oscillator strength, as well as the intervalley scattering rate.

Our $WSe_2/WS_2$ heterobilayers are formed by a stack of $WSe_2$ monolayer and $WS_2$ monolayer with a nearly 60° interlayer rotation angle. Due to the ~4% mismatch between the $WSe_2$ and $WS_2$ lattice constants, the heterobilayer can exhibit a moiré pattern with a period of ~8 nm (Fig. 1a). A carrier density $n \sim 1.8 \times 10^{12}$ cm$^{-2}$ can fill one electron per moiré cell (filling factor $v = 1$). We encapsulate the heterobilayer with hexagonal boron nitride (BN) to achieve high sample quality. The devices have both the top and bottom gates, with thin graphite as the electrodes (Fig. 1b). We measure their reflectance contrast and PL (see Methods), which reveal the absorption and emission properties of the superlattice, respectively. We have measured more than one device and present the results of the best device (Device #1) in the main paper (see Fig. S5 for the results of Devices #2 and #3). Fig. 1c displays the optical image of Device #1. It exhibits significant interlayer exciton PL even at room temperature, indicating very high device quality (Fig. 1d).

Fig. 2a-b displays the gate-dependent reflectance contrast ($\Delta R/R$) maps of Device #1 at an estimated sample temperature $T \sim 15$ K. In the charge neutrality regime near zero gate voltage, we observe two $WS_2$ intralayer moiré excitons near photon energies 1.991 and 2.082 eV (Fig. 2a) as well as three $WSe_2$ intralayer moiré excitons near 1.672, 1.724 and 1.76 eV (Fig. 2b; labeled as $X_1$, $X_2$ and $X_3$, respectively). The emergence of moiré excitons indicates the strong influence of the moiré superlattice on the excitonic states, as shown in prior studies[16, 24, 28, 29]. We inject electrons (holes) into the heterobilayer by applying equal positive (negative) gate voltage on the bottom gate ($V_{bg}$) and top gate ($V_{tg}$). From the hole to electron side, we observe spectral modulation of moiré excitons at gate voltages $V_g = V_{bg} = V_{tg}$ = -3.34, -2.02, 1.86, 3.08 V. They correspond to integer filling factors $v$ = -2, -1, 1, 2, respectively, according to our estimated charge density and comparison with prior studies[6, 7, 16, 18]. Here $v = \pm 2$ correspond to the complete filling of one moiré miniband, and $v = \pm 1$ correspond to half filling of a moiré miniband[6, 7]. The exciton spectral changes at $v = \pm 1$ signify the formation of correlated Mott phase at half filling of a miniband.

Remarkably, we also observe spectral modulation of all three WSe$_2$ moiré excitons ($X_1$, $X_2$, $X_3$) at fractional fillings $v$ = 1/3, 2/3 on the electron side. The region of interest is marked by the red dashed box in Fig. 2b. Fig. 2c displays a finer map in this region. Fig. 2d displays the second-order gate-voltage derivative of Fig. 2c to reveal the fine features. In Fig. 2d, we can observe clear spectral modulation of $X_1$, $X_2$, $X_3$ at $v$ = 1/3, 2/3. We have also examined the gate-dependent reflectance contrast profiles for $X_1$, $X_2$, $X_3$ (along the black, green, and red dashed lines in Fig. 2c). Their profiles, shown in Fig. 2e, display clear peaks or dips at $v$ = 1/3, 2/3, which signify correlated phases at these two fractional fillings.

Correlated states at $v$ = 1/3, 2/3 in WSe$_2$/WS$_2$ moiré superlattices have been reported by recent research[7, 16-19]. According to prior experiments and theoretical simulation[7, 16-21], the electron phase at $v$ = 1/3 (2/3) should correspond to generalized Wigner crystals, in which the electrons occupy 1/3 (2/3) of the moiré cells to form a periodic triangular lattice with three-fold rotational symmetry (see the illustrated charge distribution in Fig. 1e). In such electron crystalline phases, each electron is locked in its own moiré cell due to the strong Coulomb repulsion between electrons; they produce an insulating phase with no free carriers. The suppressed charge screening in these insulating phases can renormalize the bandgap and enhance the exciton binding. Therefore, we expect an increase of exciton oscillator strength and a shift of the exciton resonance energy, which are manifested as peaks or dips in the gate-dependent reflectance contrast profile, as observed in our experiment.

Compared to prior studies[7, 16-19], our results here demonstrate that fractional correlated states can significantly modulate the absorption properties of both the low- and high-lying moiré excitons. In particular, we find that these moiré excitons can exhibit very different sensitivity to the electron crystallization, which has not been shown before. As we examine their profiles in Fig. 1e, although $X_1$, $X_2$ and $X_3$ all show modulation of reflectance contrast at $v$ = 1/3, 2/3, the percentage change of reflectance contrast is much larger for $X_2$ (~54%) and $X_3$ (~18%) than $X_1$ (~1.2%). Their different behavior can be attributed to two factors. First, the high-lying $X_2$ and $X_3$ moiré excitons are more mobile than the low-lying $X_1$ excitons that are largely confined by the moiré potential. The mobile excitons are expected to be more sensitive to the larger-period structure of generalized Wigner crystals than the less mobile or confined excitons. Second, while $X_1$ mainly comes from the original A-exciton in monolayer WSe2, $X_2$, $X_3$ are generated entirely by the weak zone-folding effect of the moiré superlattice[24]. Therefore, the generalized Wigner crystal structure can have a much larger percentage impact on $X_2$, $X_3$ than $X_1$.

Besides the absorption-related signatures in the reflection measurements, we have also observed emission optical signatures of the correlated states at fractional fillings. Fig. 3a displays a gate-dependent PL map of the interlayer excitonic states in the WSe2/WS2 heterobilayer at $T \sim$ 15 K. The map shows the emission from the interlayer excitons ($IX$) and interlayer trions on the electron side ($IX^-$) and hole side ($IX^+$) (Fig. 3a). The most prominent features in the map are the drastic step-like jump of emission photon energy and intensity at integer fillings $v$ = -1 and -2 on the hole side. To reveal the fine features, we plot the gate-dependent integrated PL intensity in Fig. 3b. We observe PL enhancement features at fractional fillings $v$ = -1/4, -1/3, -8/7, -3/2 on the hole side.

The interlayer trion on the electron side also exhibits noticeable change of emission energy and intensity at $v$ = 1, 2. Compared to the hole side, the interlayer trion on the electron side exhibits many more fractional states (Fig. 3a-b). To show the fine features on the electron side, Fig. 3c displays an expanded map at $V_g = V_{bg} = V_{tg} = 0.5 – 3.5$ V, in addition to the gate-dependent integrated PL intensity profiles in Fig. 3b. From the trion energy shift and intensity enhancement, we can identify correlated states at many fractional fillings, including $v$ = 1/4, 1/3, 2/5, 2/3, 6/7, 5/4, 7/5 (4/3), 5/3, 13/7 (the filling factors as a function of $V_g$ is shown in Fig. S1). We have some ambiguity to distinguish the 7/5 and 4/3 states, whose features may overlap in our data. Similar PL maps are also observed at different positions of our device (Fig. S2). Our results are mostly consistent with recent research that measures the dielectric, conduction, and optical response of the moiré system[16, 18]. A recent arXiv preprint also reports interlayer PL signatures at fractional fillings on the electron side[17], but here we show more numerous and much finer features of the interlayer trions on both the electron and hole sides.

The numerous fractional fillings imply the existence of many correlated states with different charge orders in the moiré superlattice. Fig. 1e illustrates some possible charge order configurations for $v$ = 1/2, 2/5, 1/3, 1/4, 1/7 by using orange (silver) dots to represent moiré cells filled (unfilled) with an electron (configurations with $1 – v$ fillings can be obtained by exchanging the orange and silver color of the dots). In these phases, the electrons are locked in their own moiré cells to form a periodic structure so as to minimize their interaction potential energy. Since such crystalline electron phases are insulating, it can modulate the bandgap and the trion binding compared to the phases with mobile electrons. When a localized electron in the crystalline phase is bound with an exciton to form a trion, the trion state should also be localized in a moiré cell. Such localization may also contribute to the change of trion emission spectrum.

It is interesting to note that double trion features appear at filling factors $|v| > 1$ on both the electron and hole sides. We speculate that they arise from the following scenario. When $|v| <$ 1, each moiré cell has either zero or one existing carrier, so only the exciton or trion will be formed. But when $|v| > 1$, each moiré cell contains either one carrier or two carriers. When an exciton enters a moiré cell with one existing carrier, it will form a trion with an energy similar to the case of $|v| < 1$. When an exciton enters a moiré cell with two existing carriers, however, it will form a trion interacting with another carrier; the binding energy of this four-particle complex with respect to an exciton plus two carriers is less than the trion binding energy. This will give rise to a new emission peak at higher photon energy. As a result, two emission lines appear at $|v| > 1$. Similarly, when $|v| > 2$, an exciton can enter a moiré cell with three existing carriers to form a five-particle complex with further reduced binding; this will give rise to a new emission peak with even higher photon energy. Such a picture can qualitatively account for the emergence of new trion emission peaks at higher photon energy at $|v| = $ -2, -1, 1 in our PL data (Fig. 3a). Further research is merited to explain their quantitative behavior.

We have measured the reflection and PL maps at different temperatures, which allow us to rank the relevant energy scales in the miniband gap, Mott phases, and fractional correlated phases in the moiré superlattice (see Fig. S3, S4). The fractional filling features in both reflection and PL subside at $T > 30$ K. But the integer filling $v = $ -2 feature, which corresponds to fully filling a miniband, can sustain up to $T \sim 75$ K. The $|v| = 1$ features, which correspond to Mott insulating phases, can sustain to $T > 150$ K. These results, consistent with prior research[6, 7, 16, 18], indicate that the Mott gap is larger than the miniband energy gap, and the miniband gap is larger than the energy gap of the fractional correlated states.

Remarkably, we also observe the suppression of valley polarization of the interlayer trions at both integer and fractional fillings. Excitonic states in TMD monolayers and moiré superlattices are known to exhibit distinctive valley polarization, which is revealed by their circularly polarized luminescence[32-39]. It is interesting to study the effect of correlated phases on the valleytronic properties. Fig. 4a-b display the PL maps of the interlayer excitonic states at right-handed (R) and left-handed (L) circular polarization under the excitation of 632.8-nm laser with right-handed polarization (R). We have integrated the PL intensity ($I$) at the R-R and R-L excitation-detection polarization configurations. Fig. 4c displays the valley polarization $\eta = \frac{I_{RR} - I_{RL}}{I_{RR} + I_{RL}}$ as a function of gate voltage. Our results show a significant drop in valley polarization at $|v| = $ 1, 2. The integer-filling insulating phases can effectively enhance Coulomb interactions. As intervalley scattering is contributed by electron-hole exchange interactions, the enhancement of Coulomb interaction can facilitate the intervalley scattering and hence suppress the valley polarization[39, 40]. Remarkably, we also observe the suppression of valley polarization at $v = $ -5/3, 1/3, 2/3. These results indicate that the fractional states, with their insulating nature, can also enhance the intervalley scattering and suppress the valley polarization. Compared to the integer states, the suppression of valley polarization is weaker in the fractional states, consistent with their smaller energy gap and weaker insulation.

In summary, we have observed optical signatures of correlated states at fractional fillings in the reflectance contrast of intralayer moiré excitons, PL of the interlayer trions, as well as the interlayer trion valley polarization. The results reveal significant influence of the fractional correlated states on the absorption, emission, and valley dynamics of excitonic states in the moiré superlattice. The electron phases at fractional fillings are expected to be the stripe phases or generalized Wigner crystals. Their influence on the excitonic states, as demonstrated in our research, shall inspire researchers to further the exploration of novel excitonic and valleytronic phenomena in the regime of electron crystallization in moiré materials.

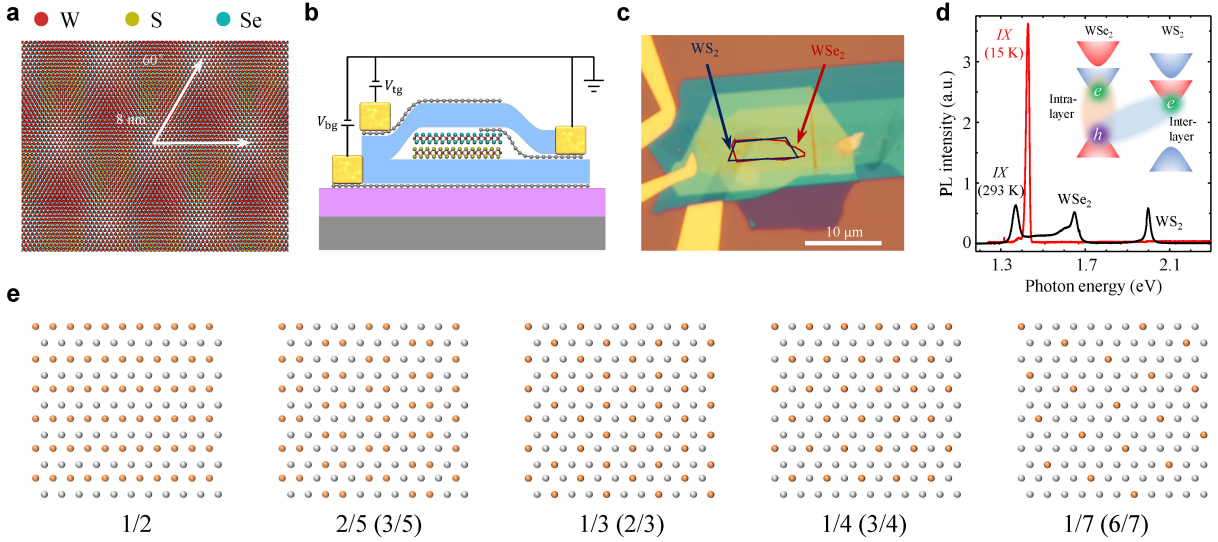

**Figure 1. WSe$_2$/WS$_2$ heterobilayer moiré superlattice and fractional correlated states. a,** Moiré pattern in the WSe$_2$/WS$_2$ heterobilayer with a 60° interlayer rotation angle. **b,** Schematic of our WSe$_2$/WS$_2$ heterobilayer devices. **c,** Optical image of Device #1. **d,** Photoluminescence (PL) spectra of the WSe$_2$/WS$_2$ heterobilayer device at sample temperature $T \sim 15$ K and $T \sim 293$ K under 532-nm continuous laser excitation. We denote the emission peaks of the intralayer excitons within the WSe$_2$ layer and WS layer and the interlayer excitons (*IX*). The inset shows the band configurations of intralayer and interlayer excitons. The band color represents different electron spins. **e,** Schematic charge distribution of possible correlated states at different fractional fillings in the moiré superlattice. The orange (silver) dots represent moiré cells filled (unfilled) with an electron. Exchanging the orange and silver color gives configurations with complementary filling factors denoted in the parentheses.

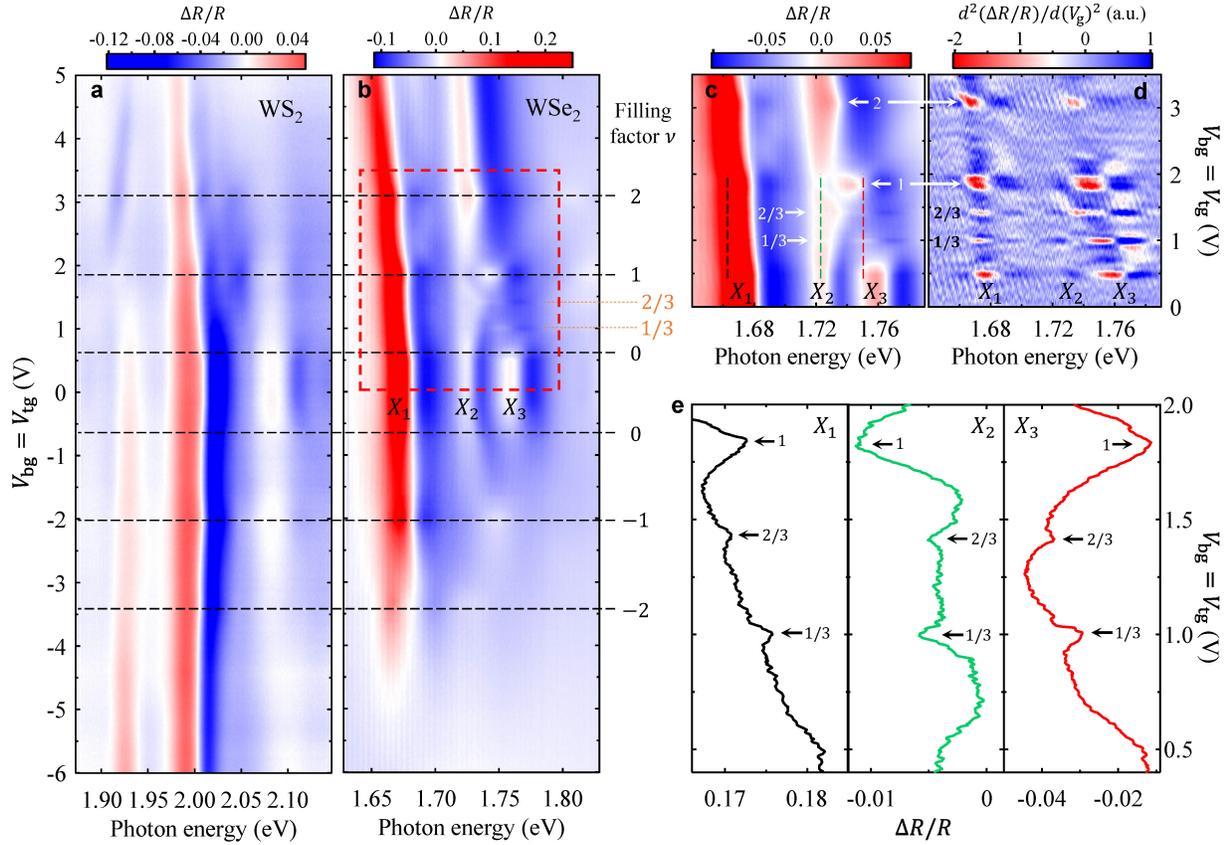

**Figure 2. Reflection optical signatures of correlated states at fractional fillings in the WSe$_2$/WS$_2$ heterobilayer moiré superlattice.** **a-b,** Gate-dependent reflectance contrast maps for the intralayer excitonic states within the WS$_2$ layer and WSe$_2$ layer in the WSe$_2$/WS$_2$ moiré superlattice. The dashed lines denote the integer and fractional filling positions. We inject electrons (holes) by applying an equal positive (negative) voltage ($V_g$) on the back gate ($V_{bg}$) and top gate ($V_{tg}$). **c,** Finer map within the red dashed box in panel **b**. **d,** Second-order gate-voltage ($V_g$) derivative of the map in panel **c**. **e,** Profiles along the three dashed lines with the corresponding color in panel **c**. Peaks and dips at integer and fractional fillings are denoted. All measurements were conducted at a sample temperature $T \sim 15$ K.

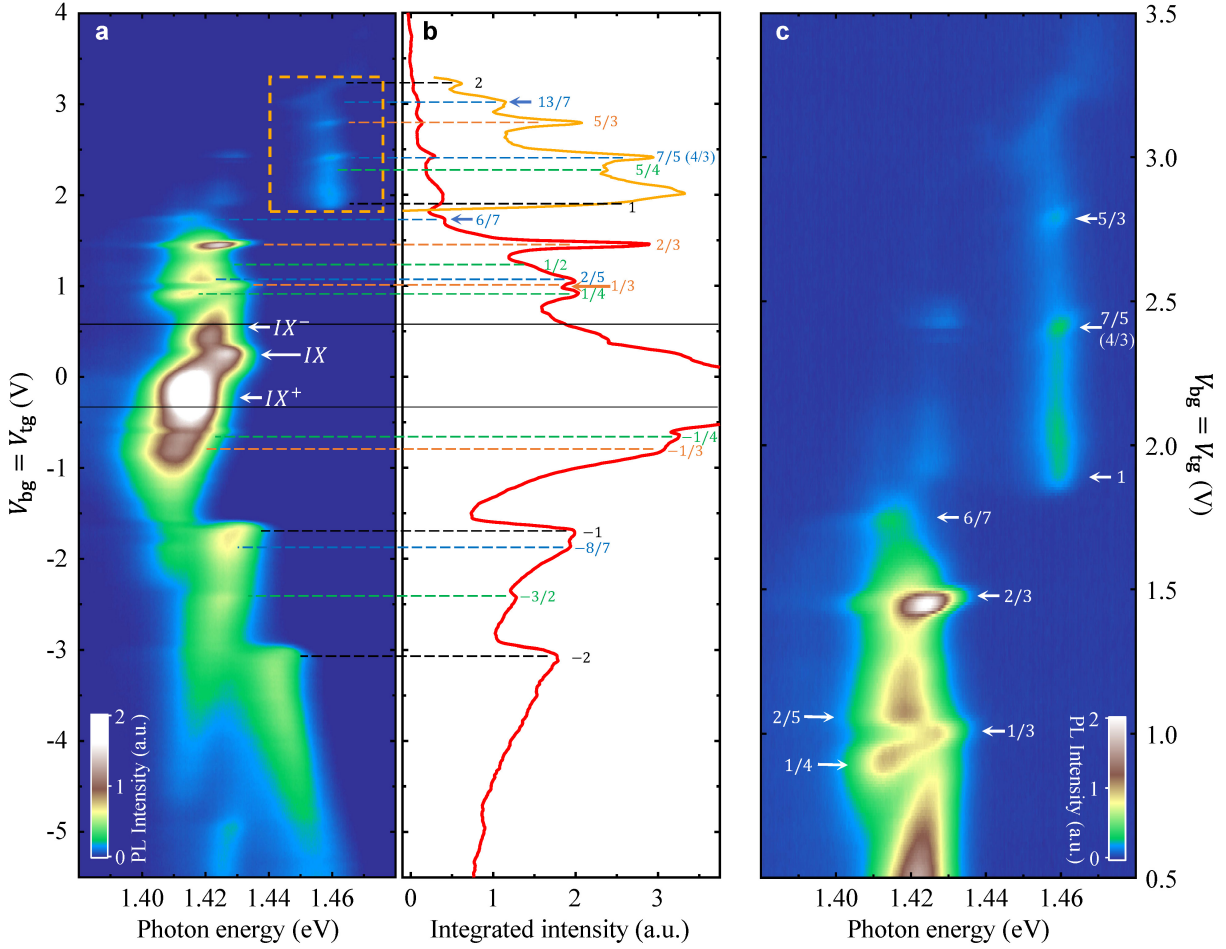

**Figure 3. Emission optical signatures of correlated states at fractional fillings in the WSe$_2$/WS$_2$ heterobilayer moiré superlattice.** **a,** Gate-dependent photoluminescence (PL) maps for the interlayer excitonic states. The PL peak in $V_g = V_{bg} = V_{tg} = 0.2 - 0.3$ V corresponds to the interlayer exciton ($IX$); all other PL features should correspond to interlayer trions or exciton-polarons ($IX^+$, $IX^-$). **b,** Integrated PL intensity in panel **a** as a function of gate voltage. The red line is the PL intensity integrated over the whole photon energy range in panel **a**; the orange line is the PL intensity integrated within the range of the orange dashed box in panel **a**. The horizontal dashed lines denote the integer and fractional filling positions. Panels **a** and **b** share the same y-axis. **c,** Zoom-in map at $V_g = 0.5 - 3.5$ V in panel **a**. The filling factors are denoted. We have some ambiguity to distinguish the 7/5 and 4/3 states, whose features may overlap in our data. The PL measurements were conducted at sample temperature $T \sim 15$ K with 532-nm continuous laser excitation at incident power of ~60 nW.

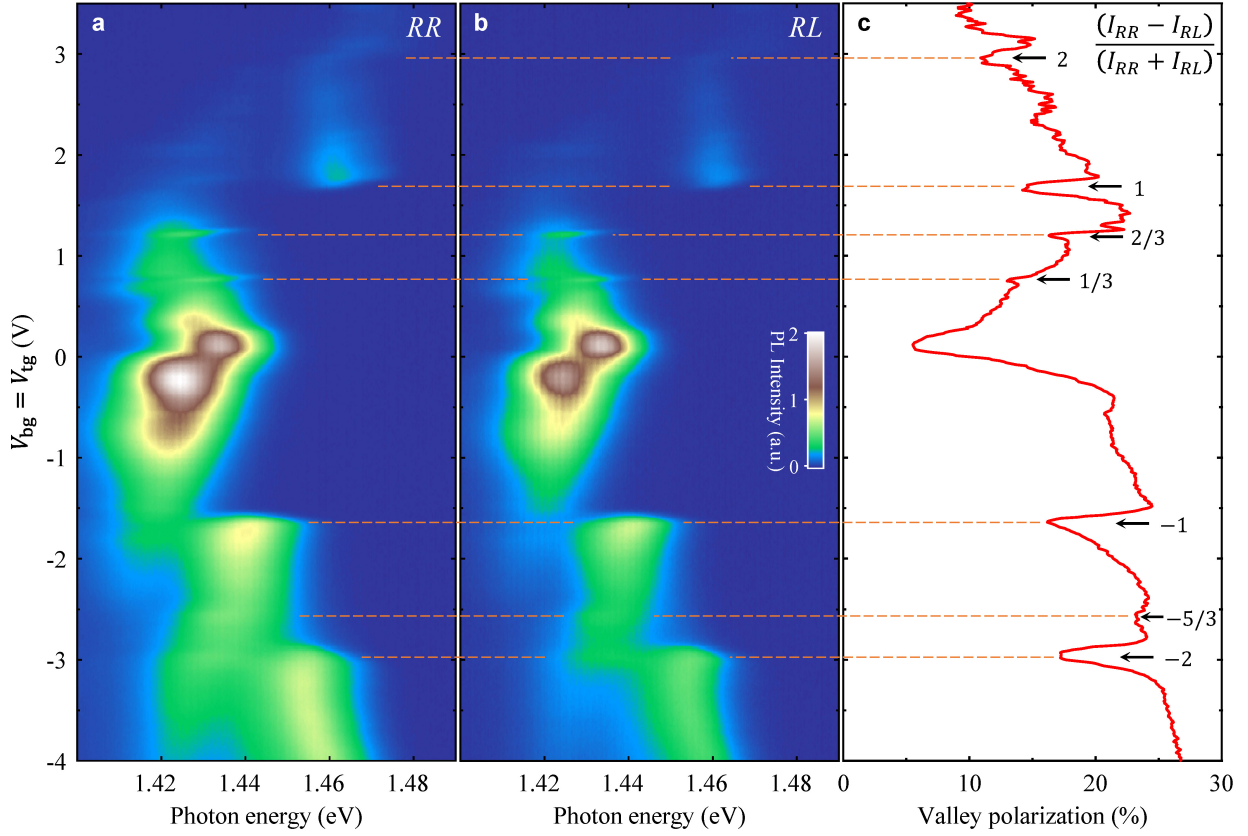

**Figure 4. Valley optical signatures of correlated states at integer and fractional fillings in the WSe$_2$/WS$_2$ heterobilayer moiré superlattice.** **a,** Gate-dependent photoluminescence (PL) maps for the interlayer excitonic states under right-handed (R) circularly polarized optical excitation and detection (RR). **b,** Similar map as panel **a** under right-handed (R) optical excitation and left-handed (L) optical detection (RL). **c,** Valley polarization as a function of gate voltage. Suppression of valley polarization at integer and fractional fillings are denoted. The PL measurements were conducted at sample temperature $T \sim 15$ K with 632.8-nm continuous laser excitation at incident power ~300 nW.

## Methods

**Device fabrication.** The WS$_2$/WSe$_2$ heterostructure devices are fabricated by the standard mechanical co-lamination of two-dimensional (2D) crystals. We use WS$_2$ and WSe$_2$ bulk crystals from HQ Graphene Inc.. The WS$_2$ and WSe$_2$ monolayers, multilayer graphene, and thin BN flakes (thickness of 15 – 35 nm) are first exfoliated onto silicon substrates with 285-nm-thick SiO$_2$ epilayer. Afterward, a polycarbonate-based dry-transfer technique is applied to stack the 2D crystals together. We first pick up thin graphite (as the top-gate electrode), and subsequently pick up a thin BN flake (as the top-gate dielectric), thin graphite (as the electrode), monolayer WS$_2$, monolayer WSe$_2$, thin BN (as the bottom-gate dielectric), and thin graphite (as the back-gate electrode). Afterward, we transfer the stack of materials onto a Si/SiO$_2$ substrate, and apply the standard electron beam lithography to pattern and deposit the gold contacts (100-nm thickness). Finally, we anneal the devices at $T$ = 350 °C for three hours in an argon environment to cleanse the interfaces.

In our preparation of the heterobilayer, we choose WS$_2$ and WSe$_2$ monolayers that have 60° (or 120°) sharp edges, and align them according to the edge direction in the stacking process. The heterobilayers thus produced are expected to have either ~0° or ~60° interlayer rotation angle. Device #1, 2, 3 all have ~60° interlayer twist angle, which can be verified by the destructive interference in the second harmonic generation. The thickness of BN flakes in the devices can be measured by atomic force microscopy (AFM). For Device #1 presented in the main paper, the thickness of both the top and bottom BN is ~25 nm.

**Low-temperature cryostat setup.** All the optical experiments were conducted in our laboratory at UC Riverside. We mount the devices inside a helium-cooled cryostat (Montana Instruments). The temperature of the cold finger in the cryostat can reach $T$ ~ 4 K. But due to the thermal insulation of the sample mount and substrate as well as the background thermal radiation, we estimate that the sample temperature is ~10 K higher than the cold-finger temperature. In this paper, all the denoted temperatures are estimated sample temperatures.

**Reflectance contrast measurements.** In the reflectance contrast measurements, we focus broadband white light onto the sample with a spot diameter of ~2 μm by an objective lens (NA = 0.6). The reflected light is collected by the same objective and analyzed by a spectrometer equipped with a charge-coupled-device (CCD) camera (IsoPlane, Princeton Instruments Inc.). To obtain the reflectance contrast, we measure a reflection spectrum on the heterobilayer sample ($R_s$) and a reference reflection spectrum ($R_r$) on a nearby area without WS$_2$ and WSe$_2$. The spectrum of reflectance contrast is obtained as $\Delta R/R = (R_s - R_r)/R_r$.

**Photoluminescence (PL) and valley polarization experiments.** For the PL experiment in Fig. 3, we use a 532-nm continuous laser (Laser Quantum; Torus 532) as the excitation light source. The laser is focused onto the sample with a spot diameter of ~1 μm through a microscope objective (NA = 0.6). The PL is collected by the same objective in backscattering geometry, and analyzed by the spectrometer with a CCD camera (IsoPlane, Princeton Instruments Inc.).

For the valley polarization PL experiment in Fig. 4, we use a helium-neon laser (wavelength 632.8

nm; Thorlabs) as the excitation light source. The linearly polarized laser passes through a quarter waveplate to obtain circular polarization. Afterward, the laser is focused onto the sample by a microscope objective (NA = 0.6). The PL signal, collected by the same objective, passes through the same quarter waveplate, which converts the right- and left-handed PL into light with two orthogonal linear polarizations. Afterward, the PL of different linear polarizations is spatially separated for ~4 mm by a calcite beam displacer (Thorlabs, BD40). The PL then passes through another quarter waveplate to become circularly polarized and enters the spectrometer. Due to their spatial separation, the two PL beams will be dispersed at different heights in the CCD camera and can be measured simultaneously. In this way, we can obtain the spectra of right- and left-handed PL simultaneously to obtain the valley polarization. This greatly facilitates our measurements of gate-dependent valley polarization.

**Estimation of carrier density and moiré period.** We calculate the density ($n$) of injected carriers at different gate voltages ($V_g$) according to the formula $ne = CV_g$, where $C = \varepsilon\varepsilon_0/d$ is the gate capacitance. Here $d$ is the BN thickness; $\varepsilon = 3.1$ is the BN dielectric constant; $\varepsilon_0$ is the permittivity of free space. In our experiments, as the top and bottom gates have nearly the same BN thickness (~ 25 nm) and capacitance, we apply equal voltage on both gates to inject carriers into the sample without inducing an electric field on it. In Fig. 3, the gate voltage spacing between two integer filling positions is about 1.36 V, which corresponds to the carrier density of about $1.87 \times 10^{12}$ cm$^{-2}$.

For one carrier per moiré unit cell ($\nu = 1$), the carrier density $n_0$ is related to the moiré period by $n_0 = 1/[L_M^2 \sin(\pi/3)]$, where $L_M = a_{\text{WSe}_2}/\sqrt{\theta^2 + \delta^2}$ is the moiré superlattice constant; $\theta$ is the twist angle; $\delta \approx 4\%$ is the mismatch between the lattice constants of WS$_2$ (0.315 nm) and WSe$_2$ (0.328 nm). For heterobilayers with exactly 0° or 60° interlayer rotation angle, the carrier density for integer filling $\nu = 1$ is $n_0 = 1.69 \times 10^{12}$ cm$^{-2}$. From the estimated carrier density $1.87 \times 10^{12}$ cm$^{-2}$ for integer filling $\nu = 1$ in Device #1, we estimate that its moiré period is 7.86 nm and the interlayer twist angle is 59.3°.

**Acknowledgments:** We thank Matthew Wilson for assistance in device fabrication, and thank Yia-Chung Chang and Kin Fai Mak for discussion. C.H.L. acknowledges support from the National Science Foundation Division of Materials Research CAREER Award No.1945660. Y.-T.C. acknowledges support from NSF under award DMR-2004701. N.M.G. acknowledges support through the Presidential Early Career Award for Scientists and Engineers (PECASE) through the Air Force Office of Scientific Research award no. FA9550-20-1-0097, and through support from the National Science Foundation Division of Materials Research CAREER award no. 1651247. K.W. and T.T. acknowledge support from the Elemental Strategy Initiative conducted by the MEXT, Japan, Grant Number JPMXP0112101001, JSPS KAKENHI Grant Number JP20H00354, and the CREST (JPMJCR15F3), JST.

**Competing interests:** The authors declare no competing interests.

**Supplementary figures:**

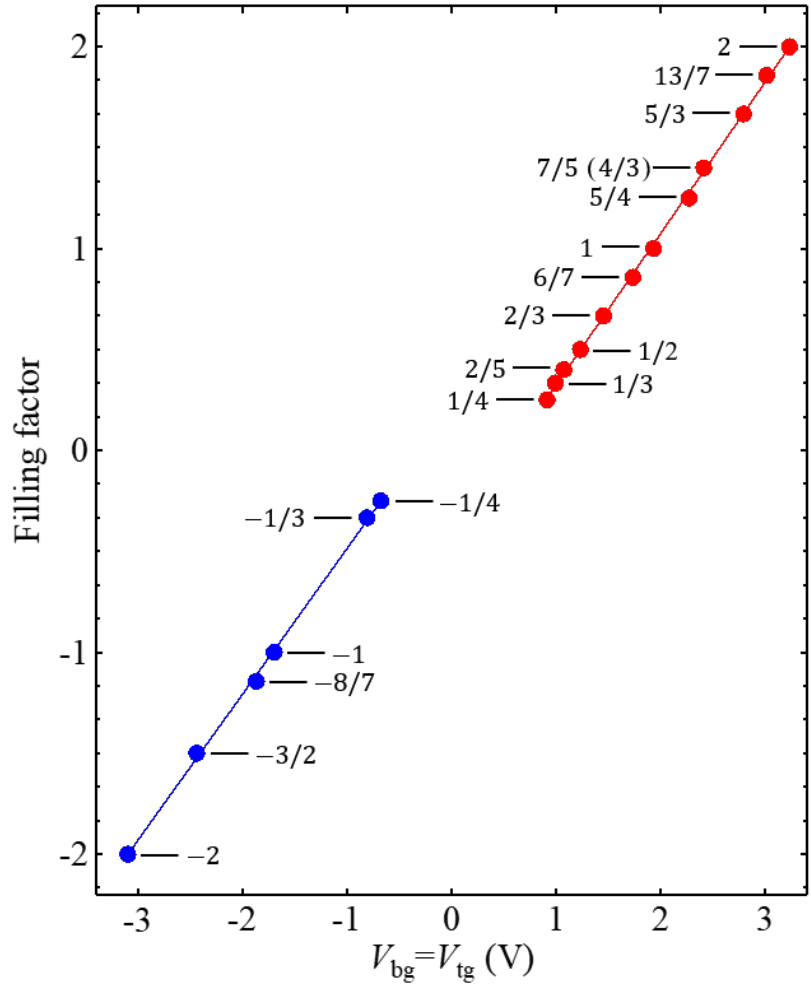

**Figure S1. The filling factors as a function of gate voltage in the PL map.** We first identify the $\nu = -2, -1, 1/3, 2/3, 1, 2$ PL features in Fig. 3a, and then use them to deduce the filling factors of other PL features. The error bar of the filling factors is within the size of the dots. Here we do linear fits on all the assigned filling factors on the electron and hole sides. The good linear fits support our assignment of the filling factors.

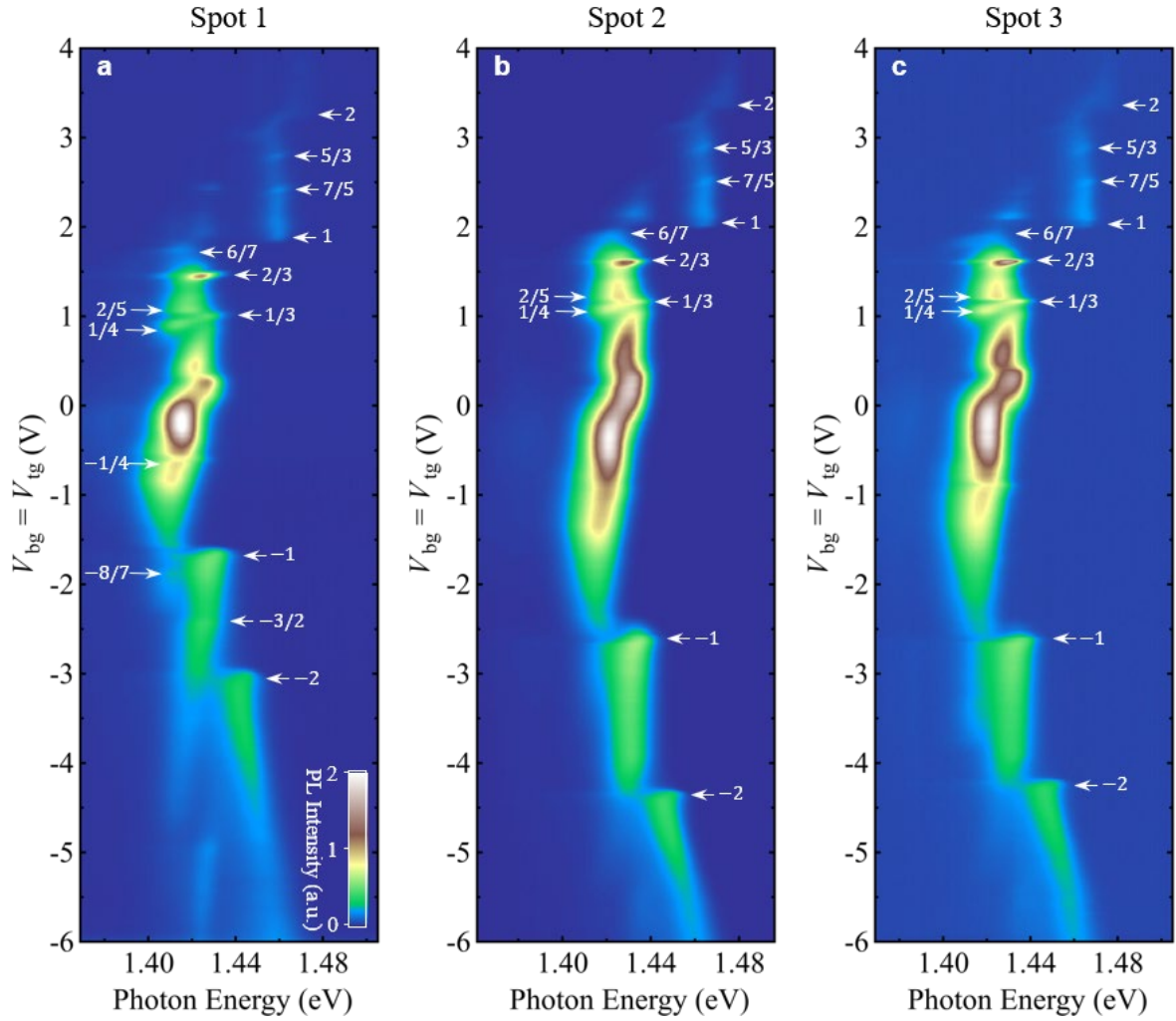

**Figure S2. Gate-dependent PL maps at different spots of Device #1. a**, **b**, **c,** The PL gate maps for Spots 1, 2, 3, respectively.   Most of the area on Device #1 shows PL signatures of fractional states.   The PL maps in Figs. 3 and 4 in the main paper are taken at Spot 1.

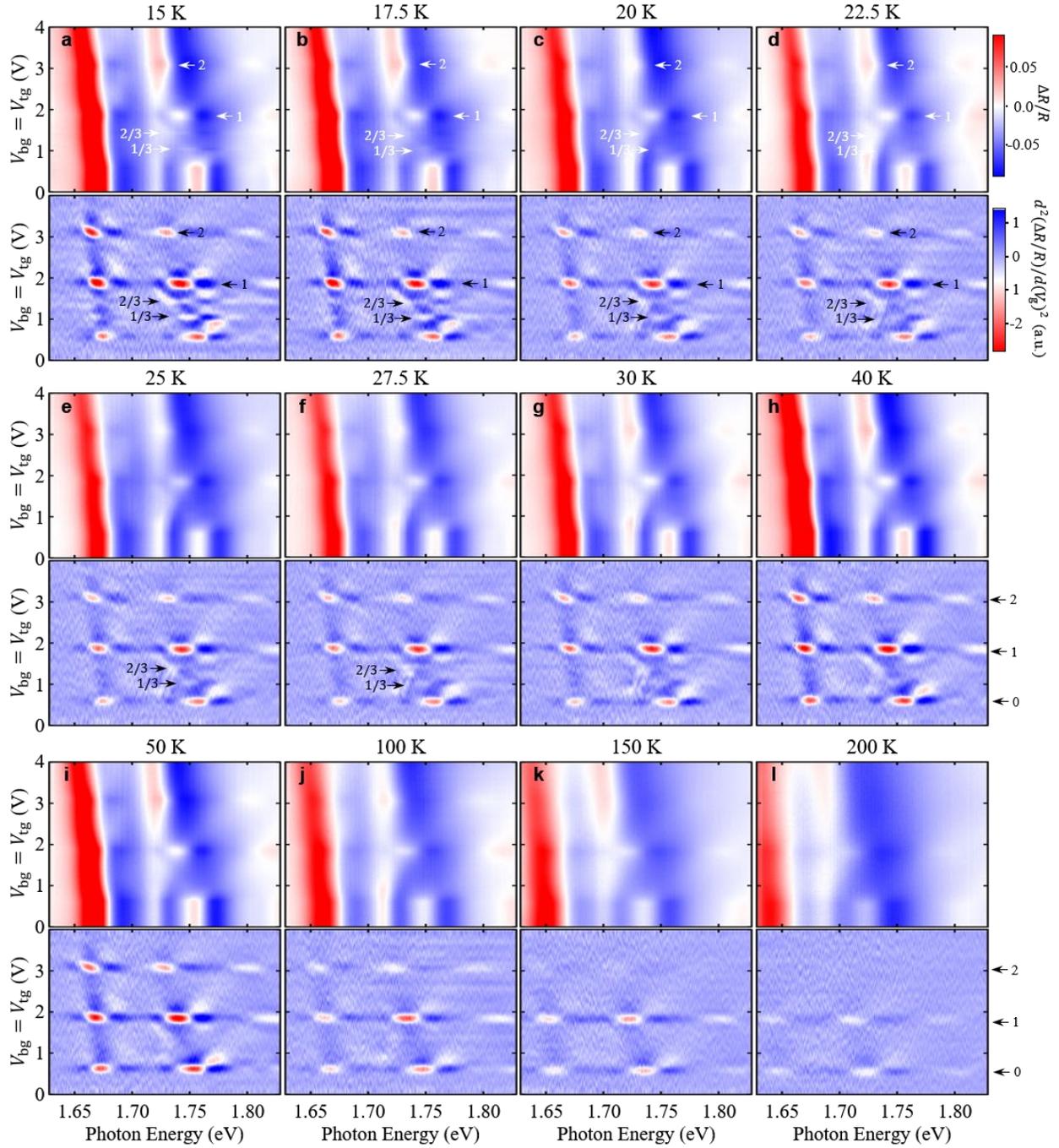

**Figure S3. Temperature-dependent reflectance contrast maps for Device #1. a-l,** The reflection contrast ($\Delta R/R$) maps (upper panels) and the second-order gate-voltage derivative of the $\Delta R/R$ maps (lower panels) at different sample temperatures from $T \sim 15$ K to $\sim 200$ K. The fractional-filling features at $\nu = 1/3, 2/3$ are observable below 30 K. The feature of Mott insulating phase at $\nu = 1$ sustains to T > 200 K.

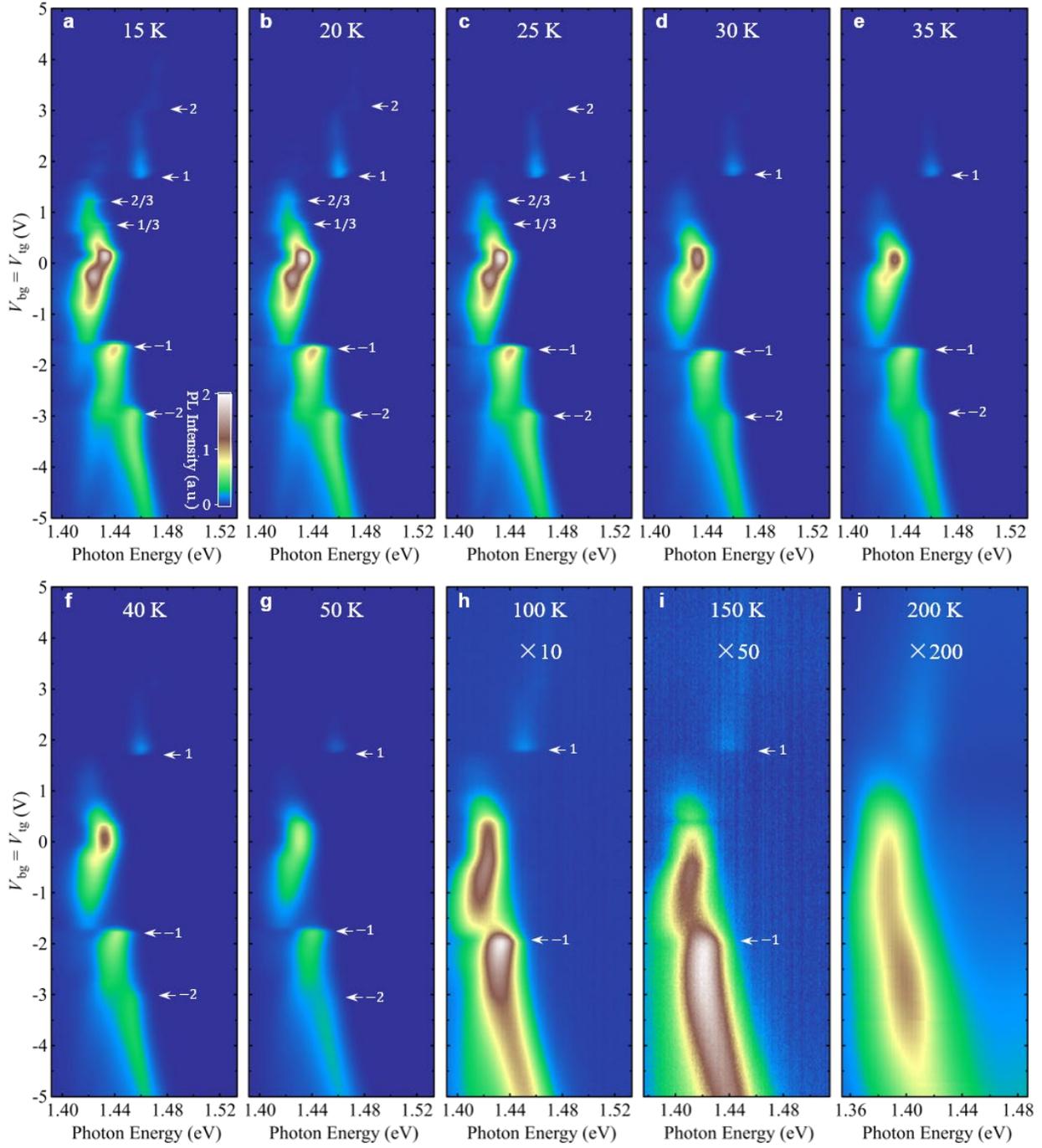

**Figure S4.  Temperature-dependent PL maps for Device #1. a-j,** The PL maps measured at different sample temperatures from ~15 K to ~200 K.   The fractional-filling features at ν = 1/3, 2/3 are observable below 30 K. The features for the Mott insulating phase at |ν| = 1 sustain to T > 150 K.

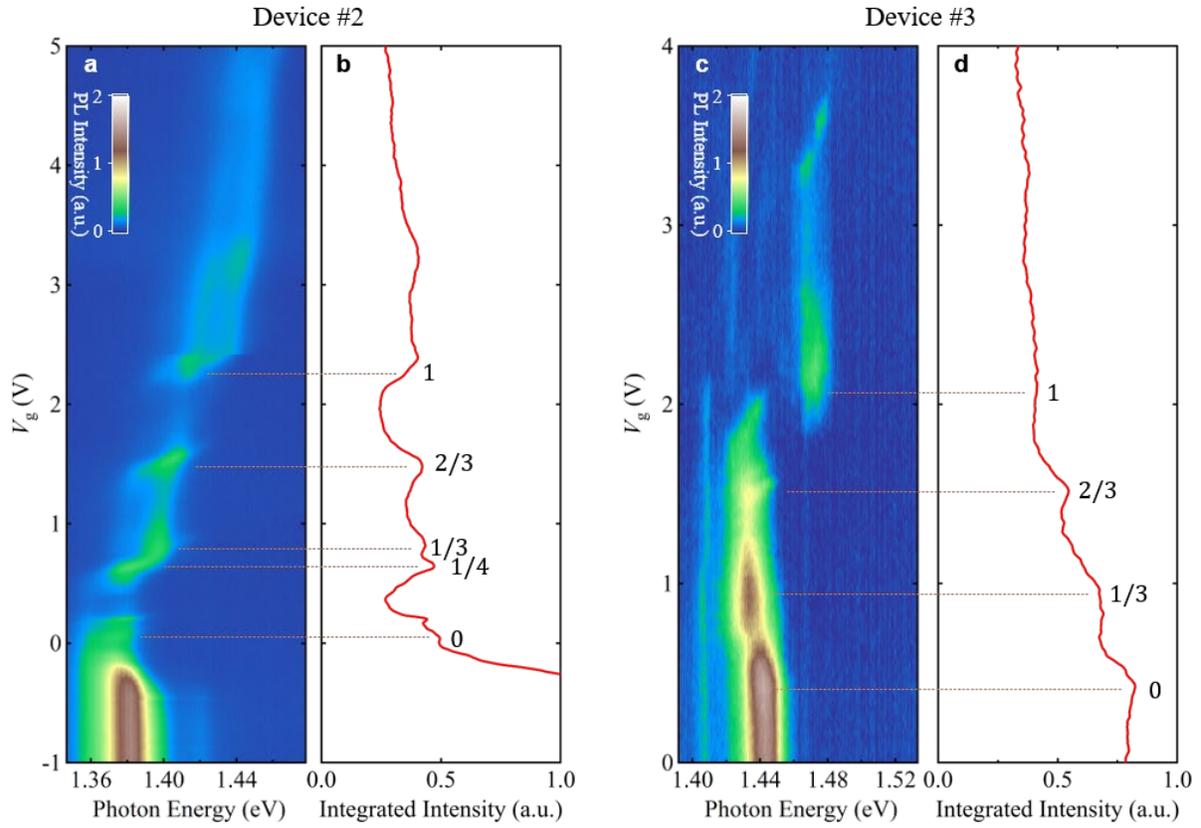

**Figure S5. Gate-dependent PL maps for two different WSe$_2$/WS$_2$ heterobilayer devices. a.** PL map of Device #2. **b**. PL map of Device #3. The observed fractional fillings are denoted in the maps.